\documentclass[runningheads,citeauthoryear]{apinv}
\usepackage{epsfig,cite,graphics}
\usepackage[T2A]{fontenc}
\usepackage[cp1251]{inputenc}
\usepackage[bulgarian,english]{babel}

\begin{document}

\title{NPM1G\,--10.0586: an emission-line companion of the Seyfert galaxy Mrk\,509}
\titlerunning{NPM1G\,--10.0586: an emission-line galaxy}
\author{Lyuba Slavcheva-Mihova, Boyko Mihov}
\authorrunning{L. Slavcheva-Mihova, B. Mihov}
\tocauthor{L. Slavcheva-Mihova, B. Mihov}
\institute{Institute of Astronomy and NAO, Bulgarian Academy of Sciences, BG-1784 Sofia\newline
\email{lslav@astro.bas.bg}}
\papertype{Conference poster. Accepted on 20.01.2012}	
\maketitle

\begin{abstract}
We report spectral observations of the galaxy NPM1G\,--10.0586, the main
candidate-companion of Mrk\,509. Mrk\,509 is a Seyfert galaxy showing no evidence of morphological
perturbations of the potential. The spectrum of NPM1G\,--10.0586 obtained by us is emission-line. The
derived weighted mean redshift is $0.03313\pm0.00023$, which makes NPM1G\,--10.0586 a physical companion of Mrk\,509.
\end{abstract}



\section*{Introduction}
Simulations show that interactions could induce tails, bridges, and asymmet\-ries. The relation between galaxy interactions and the onset
of nuclear activity is founded upon the key studies of Toomre \& Toomre
(1972) and Gunn (1979). Tidal interactions could lead to gas inflow (Byrd
et al. 1986; Noguchi 1988) and thus to circumnuclear star formation or even
fueling the active nucleus.

With an absolute magnitude of $M{_B}$\,=\,$-22\fm5$ (V\'eron-Cetty \& V\'eron 2010),
Mrk\,509 lies close to the boundary in luminosity between Seyfert\,1 nuclei and quasars.
It is classified as SA0 (Slavcheva-Mihova \& Mihov 2011a) with a global ellipticity
of 0.16 (Slavcheva-Mihova \& Mihov 2011b). The former paper analyses the
presence of bars, rings, light asymmetries, and companions in a sample of
35 Seyfert galaxies and in a control sample of inactive galaxies. Mrk\,509 is
among the three Seyfert galaxies with none of the above features present.
Although there is no direct evidence of a companion, such a possibility was
hinted by Rafanelli et al. (1993) and Boris et al. (2002).

With the aim to clarify the local environment of Mrk\,509, we obtained
spectra of the galaxy NPM1G\,--10.0586, the main candidate-companion of
Mrk\,509.

 \begin{figure}[t]
 \centering{\epsfig{file=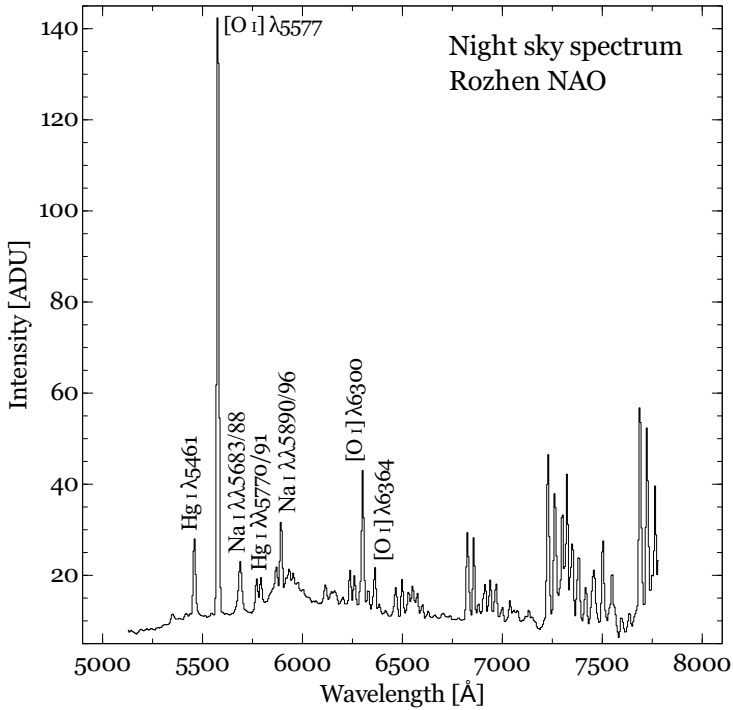, width=0.75\textwidth}}
 \caption[]{Night sky spectrum at a mean zenith distance of 65\degr.}
 \label{sky_sp}
 \end{figure}
 \begin{figure}[t]
 \centering{\epsfig{file=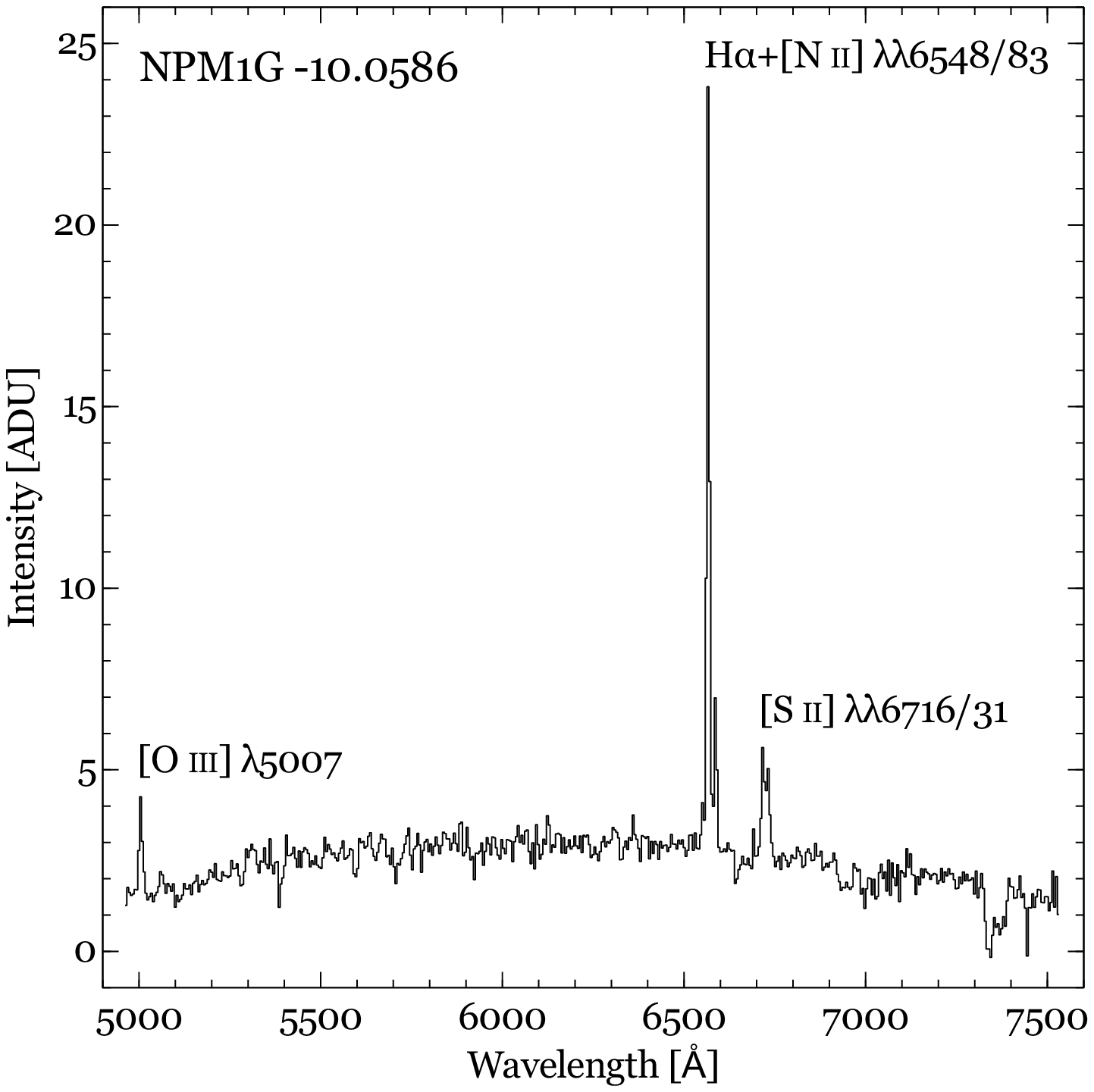, width=0.75\textwidth}}
 \caption[]{De-redshifted spectrum of the galaxy NPM1G\,--10.0586.}
 \label{obj_sp}
 \end{figure}

\section{Spectroscopy}
The spectral observations of the galaxy NPM1G\,--10.0586 were performed on
August 7$^{\rm th}$, 2011 with the 2-m telescope of the Rozhen National Astronomical
Observatory, equipped with a two-channel focal reducer. We used a grism
with 300 grooves/mm and 512$\times$512 Princeton Instruments VersArray:512B
CCD detector with a square pixel size of 24 $\mu$m (a scale of 0\farcs884/px).
This observational setup gives a spectral resolution of 5.15 \AA/px in the range
4800--7500 \AA. Several spectra of the target were acquired with an exposure
time of 600 sec each. Bias and flat field frames were taken after the target
frames; a quartz lamp was used as a flat field source. No spectrophotometric
standard was observed due to the high airmass of the target, which
varied from 1.89 to 2.93 during the observations.
The weighted mean full width at half maximum of the stellar images is 1\farcs39.

The reduction of the individual spectra was performed using IDL\,7 and consists of bias subtraction,
flat fielding, cosmic ray hit cleaning (L.A.COS\-MIC, van Dokkum 2001),
sky subtraction, and one-dimensional spectrum extraction. The spectra with highest signal-to-noise ratio
were then averaged. The wavelength calibration was done using
the following night sky lines (Fig.~\ref{sky_sp}): [O\,$\rm \scriptstyle I$] $\lambda\lambda$5577.339, 6300.304, 6363.776
(resulting from natural sky light) and Hg\,$\rm \scriptstyle I$ $\lambda\lambda$5460.754, 5769.598, 5790.663 (owing to light pollution). 
The Hg I lines are strong because at the target position on the sky the contribution from the Pamporovo resort lighting is considerable. 
The accuracy of the wavelength calibration was estimated to be $\pm$1.76 \AA. No flux calibration was done. 

We identified the following
emission lines in the spectrum of the galaxy NPM1G\,--10.0586: [O\,$\rm \scriptstyle III$] $\lambda$5006.843,
[N\,$\rm \scriptstyle II$] $\lambda\lambda$6548.05, 6583.45, $\rm H\alpha$ $\lambda$6562.801,
and [S\,$\rm \scriptstyle II$] $\lambda\lambda$6716.44, 6730.82. The weighted mean redshift, derived from these lines, is
$0.03313\pm0.00023$. The final de-redshifted spectrum of the galaxy NPM1G\,--10.0586 is shown in Fig.~\ref{obj_sp}.

\section{Discussion}
Mrk\,509 ($z$\,=\,$0.034397\pm0.000040$, Fisher et al. 1995) is one of the few Seyfert
galaxies in the sample of Slavcheva-Mihova \& Mihov (2011a) with no evidence
of morphological perturbations of the potential. We performed spectral obser\-vations
of NPM1G\,--10.0586 with the aim to find out whether it is a companion of
Mrk\,509. We obtained an emission-line spectrum.
We identified the following emission lines:
[O\,$\rm \scriptstyle III$] $\lambda$5006.843,
[N\,$\rm \scriptstyle II$] $\lambda\lambda$6548.05, 6583.45, $\rm H\alpha$ $\lambda$6562.801,
and [S\,$\rm \scriptstyle II$] $\lambda\lambda$6716.44, 6730.82.
The resulting weighted mean redshift is $0.03313\pm0.00023$. This makes
NPM1G\,--10.0586 an {\em  emission-line physical companion} of the Seyfert galaxy Mrk\,509.

\section*{Acknowledgements}
The Two-Channel Focal Reducer was transferred
to the Rozhen National Astronomical Observatory under a contract between the
Institute of Astrono\-my and National Astronomical Observatory, Bulgarian
Academy of Sciences, and the Max Planck Institute for Solar System Research.


\begin{thebibliography}{}

\bibitem{}
Boris N. V., Donzelli C. J., Pastoriza M. G., Rodriguez-Ardila A., Ferreiro D. L., 2002, {\em A\&A} 384, 780

\bibitem{}
Byrd G., Valtonen M., Sundelius B., Valtaoja L., 1986, {\em A\&A} 166, 75

\bibitem{}
Fisher K. B., Huchra J. P., Strauss M. A., Davis M., Yahil A., Schlegel D., 1995, {\em ApJS} 100, 69

\bibitem{}
Gunn J., 1979, Active Galactic Nuclei, eds. C. Hazard \& S. Mitton (Cambridge University Press, Cambridge), 213

\bibitem{}
Noguchi M., 1988, {\em A\&A} 203, 259

\bibitem{}
Rafanelli P., Marziani P., Birkle K., Thiele U., 1993, {\em A\&A} 275, 451

\bibitem{}
Slavcheva-Mihova L., Mihov B., 2011a, {\em A\&A} 526, A43

\bibitem{}
Slavcheva-Mihova L., Mihov B., 2011b, {\em AN} 332, 191

\bibitem{}
Toomre A., Toomre J., 1972, {\em ApJ} 178, 623 

\bibitem{}
van Dokkum P. G., 2001, {\em PASP} 113, 1420

\bibitem{}
V\'eron-Cetty M.-P., V\'eron P., 2010, {\em A\&A} 518, A10

\end{thebibliography}
\end{document}